\documentclass{aastex63}

\graphicspath{./}
\DeclareUnicodeCharacter{0301}{\'{i}}
\DeclareUnicodeCharacter{0300}{\'{i}}

\submitjournal{The Astrophysical Journal Letters}
\received{March 7, 2022}
\revised{May 26, 2022}
\accepted{May 26, 2022}

\shorttitle{MUSSES2020J}
\shortauthors{J. Jiang et al.}

\begin{document}

\title{MUSSES2020J: The Earliest Discovery of a Fast Blue Ultraluminous Transient at Redshift 1.063}

\correspondingauthor{Ji-an Jiang}
\email{jian.jiang@nao.ac.jp}

\author[0000-0002-9092-0593]{Ji-an Jiang}
\affiliation{Kavli Institute for the Physics and Mathematics of the Universe (WPI), The University of Tokyo Institutes for Advanced Study, The University of Tokyo, 5-1-5 Kashiwanoha, Kashiwa, Chiba 277-8583, Japan}
\affiliation{National Astronomical Observatory of Japan, National Institutes of Natural Sciences, 2-21-1 Osawa, Mitaka, Tokyo 181-8588, Japan}

\author{Naoki Yasuda}
\affiliation{Kavli Institute for the Physics and Mathematics of the Universe (WPI), The University of Tokyo Institutes for Advanced Study, The University of Tokyo, 5-1-5 Kashiwanoha, Kashiwa, Chiba 277-8583, Japan}

\author[0000-0003-2611-7269]{Keiichi Maeda}
\affiliation{Department of Astronomy, Kyoto University, Kitashirakawa-Oiwake-cho, Sakyo-ku, Kyoto 606-8502, Japan}

\author[0000-0001-8537-3153]{Nozomu Tominaga}
\affiliation{National Astronomical Observatory of Japan, National Institutes of Natural Sciences, 2-21-1 Osawa, Mitaka, Tokyo 181-8588, Japan}
\affiliation{Department of Physics, Faculty of Science and Engineering, Konan University, 8-9-1 Okamoto, Kobe, Hyogo 658-8501, Japan}
\affiliation{Kavli Institute for the Physics and Mathematics of the Universe (WPI), The University of Tokyo Institutes for Advanced Study, The University of Tokyo, 5-1-5 Kashiwanoha, Kashiwa, Chiba 277-8583, Japan}

\author{Mamoru Doi}
\affiliation{Institute of Astronomy, Graduate School of Science, The University of Tokyo, 2-21-1 Osawa, Mitaka, Tokyo 181-0015, Japan}
\affiliation{Research Center for the Early Universe, Graduate School of Science, The University of Tokyo, 7-3-1 Hongo, Bunkyo-ku, Tokyo 113-0033, Japan}
\affiliation{Kavli Institute for the Physics and Mathematics of the Universe (WPI), The University of Tokyo Institutes for Advanced Study, The University of Tokyo, 5-1-5 Kashiwanoha, Kashiwa, Chiba 277-8583, Japan}

\author[0000-0001-5250-2633]{\v{Z}eljko Ivezi\'{c}}
\affiliation{Department of Astronomy, University of Washington, Box 351580, Seattle, Washington 98195-1580, USA}

\author[0000-0003-2874-6464]{Peter Yoachim}
\affiliation{Department of Astronomy, University of Washington, Box 351580, Seattle, Washington 98195-1580, USA}

\author[0000-0002-6765-8988]{Kohki Uno}
\affiliation{Department of Astronomy, Kyoto University, Kitashirakawa-Oiwake-cho, Sakyo-ku, Kyoto 606-8502, Japan}

\author[0000-0003-1169-1954]{Takashi J. Moriya}
\affiliation{National Astronomical Observatory of Japan, National Institutes of Natural Sciences, 2-21-1 Osawa, Mitaka, Tokyo 181-8588, Japan}
\affiliation{School of Physics and Astronomy, Faculty of Science, Monash University, Clayton, VIC 3800, Australia}

\author[0000-0001-7225-2475]{Brajesh Kumar}
\affiliation{Aryabhatta Research Institute of Observational Sciences, Manora Peak, Nainital, 263 001 India}

\author[0000-0001-8415-6720]{Yen-Chen Pan}
\affiliation{Graduate Institute of Astronomy, National Central University, 300 Jhongda Road, Zhongli, Taoyuan, 32001, Taiwan}

\author[0000-0002-5011-5178]{Masayuki Tanaka}
\affiliation{National Astronomical Observatory of Japan, National Institutes of Natural Sciences, 2-21-1 Osawa, Mitaka, Tokyo 181-8588, Japan}

\author[0000-0001-8253-6850]{Masaomi Tanaka}
\affiliation{Astronomical Institute, Tohoku University, Aoba, Sendai 980-8578, Japan}
\affiliation{Kavli Institute for the Physics and Mathematics of the Universe (WPI), The University of Tokyo Institutes for Advanced Study, The University of Tokyo, 5-1-5 Kashiwanoha, Kashiwa, Chiba 277-8583, Japan}

\author[0000-0001-9553-0685]{Ken'ichi Nomoto}
\affiliation{Kavli Institute for the Physics and Mathematics of the Universe (WPI), The University of Tokyo Institutes for Advanced Study, The University of Tokyo, 5-1-5 Kashiwanoha, Kashiwa, Chiba 277-8583, Japan}

\author[0000-0001-8738-6011]{Saurabh W. Jha}
\affiliation{Department of Physics and Astronomy, Rutgers, The State University of New Jersey, 136 Frelinghuysen Road, Piscataway, New Jersey 08854, USA}

\author[0000-0001-9046-4420]{Pilar Ruiz-Lapuente}
\affiliation{Instituto de Física Fundamental, Consejo Superior de Investigaciones Científicas, Calle de Serrano 121, E-28006 Madrid, Spain}
\affiliation{Institut de Ciències del Cosmos (UB-IEEC), Calle de Martí i Franqués 1, E-08028 Barcelona, Spain}

\author[0000-0003-3947-5946]{David Jones}
\affiliation{Instituto de Astrof\'isica de Canarias, E-38205 La Laguna, Spain}
\affiliation{Departamento de Astrof\'isica, Universidad de La Laguna, E-38206 La Laguna, Spain}

\author[0000-0002-4060-5931]{Toshikazu Shigeyama}
\affiliation{RESCEU, Graduate School of Science, The University of Tokyo, 2-21-1 Osawa, Mitaka, Tokyo 181-0015, Japan}

\author[0000-0001-7266-930X]{Nao Suzuki}
\affiliation{Kavli Institute for the Physics and Mathematics of the Universe (WPI), The University of Tokyo Institutes for Advanced Study, The University of Tokyo, 5-1-5 Kashiwanoha, Kashiwa, Chiba 277-8583, Japan}
\affiliation{Lawrence Berkeley National Lab, Berkeley, CA 94720, USA}

\author[0000-0001-6402-1415]{Mitsuru Kokubo}
\affiliation{Department of Astrophysical Sciences, Princeton University, Princeton, New Jersey 08544,USA}

\author[0000-0002-6174-8165]{Hisanori Furusawa}
\affiliation{National Astronomical Observatory of Japan, National Institutes of Natural Sciences, 2-21-1 Osawa, Mitaka, Tokyo 181-8588, Japan}

\author{Satoshi Miyazaki}
\affiliation{National Astronomical Observatory of Japan, National Institutes of Natural Sciences, 2-21-1 Osawa, Mitaka, Tokyo 181-8588, Japan}
\affiliation{SOKENDAI (The Graduate University for Advanced Studies), Mitaka, Tokyo 181-8588, Japan}

\author[0000-0001-5576-8189]{Andrew J. Connolly}
\affiliation{Department of Astronomy, University of Washington, Box 351580, Seattle, Washington 98195-1580, USA}

\author[0000-0002-6688-0800]{D. K. Sahu}
\affiliation{Indian Institute of Astrophysics, II Block Koramangala,Bangalore 560034, India}

\author[0000-0003-3533-7183]{G. C. Anupama}
\affiliation{Indian Institute of Astrophysics, II Block Koramangala,Bangalore 560034, India}




\vspace{10pt}

\begin{abstract}

In this Letter, we report the discovery of an ultraluminous fast-evolving transient in rest-frame UV wavelengths, MUSSES2020J, soon after its occurrence by using the Hyper Suprime-Cam (HSC) mounted on the 8.2 m Subaru telescope. The rise time of about 5 days with an extremely high UV peak luminosity shares similarities to a handful of fast blue optical transients whose peak luminosities are comparable with the most luminous supernovae while their timescales are significantly shorter (hereafter ``fast blue ultraluminous transient," FBUT). In addition, MUSSES2020J is located near the center of a normal low-mass galaxy at a redshift of 1.063, suggesting a possible connection between the energy source of MUSSES2020J and the central part of the host galaxy. Possible physical mechanisms of powering this extreme transient such as a wind-driven tidal disruption event and an interaction between supernova and circumstellar material are qualitatively discussed based on the first multiband early-phase light curve of FBUTs, although whether the scenarios can quantitatively explain the early photometric behavior of MUSSES2020J requires systematical theoretical investigations. Thanks to the ultrahigh luminosity in UV and blue optical wavelengths of these extreme transients, a promising number of FBUTs from the local to the high-$z$ universe can be discovered through deep wide-field optical surveys in the near future.

\end{abstract}


\section{Introduction} \label{sec:intro}

With the booming development of wide-field survey facilities and the growth of high-cadence time-domain surveys, studies of a vast range of extragalactic transients from gamma-ray to radio wavelengths have been widely carried out in recent years. The fast blue optical transient (FBOT) is a new transient type confirmed several years ago and subsequently investigated by several transient survey projects recently \citep{drout14,tanaka16,pursiainen18,tampo20,ho21}. The majority of FBOTs reach peak brightness of $-17$ to $-20$ mag in blue optical wavelengths within 10 days after the discovery. Although the origin of these peculiar objects is still under debate, the rapid evolution may be explained by an explosion of a stellar progenitor with an ultra-stripped envelope \citep{suwa15,moriya17,tolstov19}.

In 2018, the discovery of an extremely luminous and fast-evolving FBOT, AT~2018cow, challenged previous hypotheses as an unfeasible extended and low-mass envelope would be required to explain such an unprecedented object \citep{prentice18,perley19}. Therefore, new mechanisms such as electron-capture collapse \citep{lyutikov19}, a tidal disruption event (TDE) induced by an intermediate-mass black hole (IMBH; \citealp{perley19,kuin19}), a common envelope jet \citep{soker19,soker22}, magnetar formation \citep{mohan20}, fallback accretion following the collapse of a massive star \citep{margutti19}, circumstellar material (CSM) interaction of a pulsational pair-instability supernova \citep{leung20b, leung21}, and cooling emission powered by a shocked jet \citep{gottlieb22} have been proposed while the nature of such kinds of extreme FBOTs is still under active debate. Although dozens of FBOT events have been reported in recent years, only a handful of them show a peak absolute magnitude above $-20$ mag with a rest-frame rise time within about one week in optical/UV wavelengths and were discovered at relatively high redshifts \citep{vinko15,ho20,perley21}, indicating the rarity of this extreme transient type. Hereafter, we suggest calling this type of transient as a ``fast blue ultraluminous transient" (``FBUT").

Recently, a statistical study of fast transients was conducted by \citet{ho21} via the Zwicky Transient Facility (ZTF; \citealp{ZTF19a,ZTF19b,ZTF19c}). Their sample selection is based on transient light curves at the bright phase (i.e., the time spent above half-maximum brightness of the light curve) because the shallow imaging depth ($\lesssim 21$ mag in $r$ band) makes it hard for the survey to observe the fast transients in the early phase. Thanks to the deep and wide imaging capability of Subaru Hyper Suprime-Cam (HSC), we are able to discover transients from an extremely early/faint phase at a large distance range with typical $g$-band observations down to 26 mag (5-$\sigma$), making Subaru/HSC the most powerful facility for studying the rising behavior of various fast transients before the upcoming Vera C. Rubin Observatory Legacy Survey of Space and Time (Rubin/LSST; \citealp{ivezic19}). 

Since 2014, a series of transient surveys aiming to discover supernovae (SNe) within a few days of their explosions have been carried out through Subaru open-use programs \citep{tanaka16,JJA2017,tominaga19}. The ``MUlti-band Subaru Survey for Early-phase Supernovae" (MUSSES) is a representative project which is designed to catch the first light of SNe with Subaru/HSC. Here we report the earliest discovery of an FBUT, MUSSES2020J, during the MUSSES observing campaign in 2020 December. This Letter is structured as follows. An introduction to the MUSSES2020 campaign and the observations of MUSSES2020J is presented in Section 2. The characteristics of MUSSES2020J and its host galaxy are introduced in Sections 3 and 4, respectively. Further discussion on the possible energy source and early photometric behavior of MUSSES2020J and our conclusions are given in Section 5. Throughout the paper we adopt the flat $\Lambda$CDM cosmology with Hubble constant $H_{0}$ = 70 km s$^{-1}$ Mpc$^{-1}$ and $\Omega_{M}=0.3$. All magnitudes are given in the AB system.

\section{Observations of MUSSES2020J}

\subsection{The MUSSES2020 Campaign}
The MUSSES project was proposed in 2014. The major scientific goal of MUSSES is to identify the progenitor and explosion mechanism of Type Ia supernovae (SNe Ia) by using multiband photometric information soon after SN explosions. The MUSSES project is composed of two parts, the Subaru/HSC multiband survey and the imaging/spectroscopic follow-ups with 1--10m class telescopes via the MUSSES global follow-up network. The formal observation of MUSSES was started in 2016 \citep{JJA2017}. The third MUSSES observing campaign (``MUSSES2020") which includes 10 consecutive half-night Subaru/HSC observations (scheduled on UT 2020 December 11 to 20) was carried out from 2020 December to 2021 March. The survey monitored a $\sim$31 deg$^2$ sky region (19 HSC pointings with a few percentage overlaps) selected from the Hyper Suprime-Cam Subaru Strategic Program (HSC SSP; \citealp{aihara18,miyazaki18}) in $g$ and $r$ bands. As in previous MUSSES campaigns, we adopt different observing depths during the survey in order to make the best use of telescope time to study early-phase SNe. Standard image reduction procedures including bias, dark, flat, and fringe corrections, as well as astrometric and photometric calibrations against the Pan-STARRS1 (PS1) $3\pi$ catalog \citep{tonry12,magnier13} were done with the HSC pipeline, a version of the LSST stack \citep{axelrod10,bosch18,ivezic19}. Image subtraction was applied using deep, coadded template images created from data taken by HSC SSP since 2014. The forced point-spread function photometry was performed at the positions of transients left in the template-subtracted images. We refer to \citet{aihara18} and \citet{yasuda19} for further details on the HSC data reduction.

\subsection{Discovery and Follow-ups of MUSSES2020J}

Although the MUSSES2020 Subaru/HSC survey was performed under poor weather conditions (with a survey completeness of $\sim$33\%), 20 transients that show fast-rising behavior were discovered\footnote{Details of other MUSSES2020 fast transients will be given in separate papers.}. Out of the 20 fast transients, an ultraluminous fast transient at host photometric redshift of $\sim$1 was discovered on UT December 11.31, the first night of the MUSSES2020 campaign. We designated this transient as MUSSES2020J (AT~2020afay), the 10th fast transient discovered in the MUSSES2020 campaign. From the Gamma-ray Coordinates Network circulars archive, no gamma-ray burst alert near the position of MUSSES2020J was reported before and during the MUSSES2020 HSC survey.

\begin{figure*}
\gridline{\fig{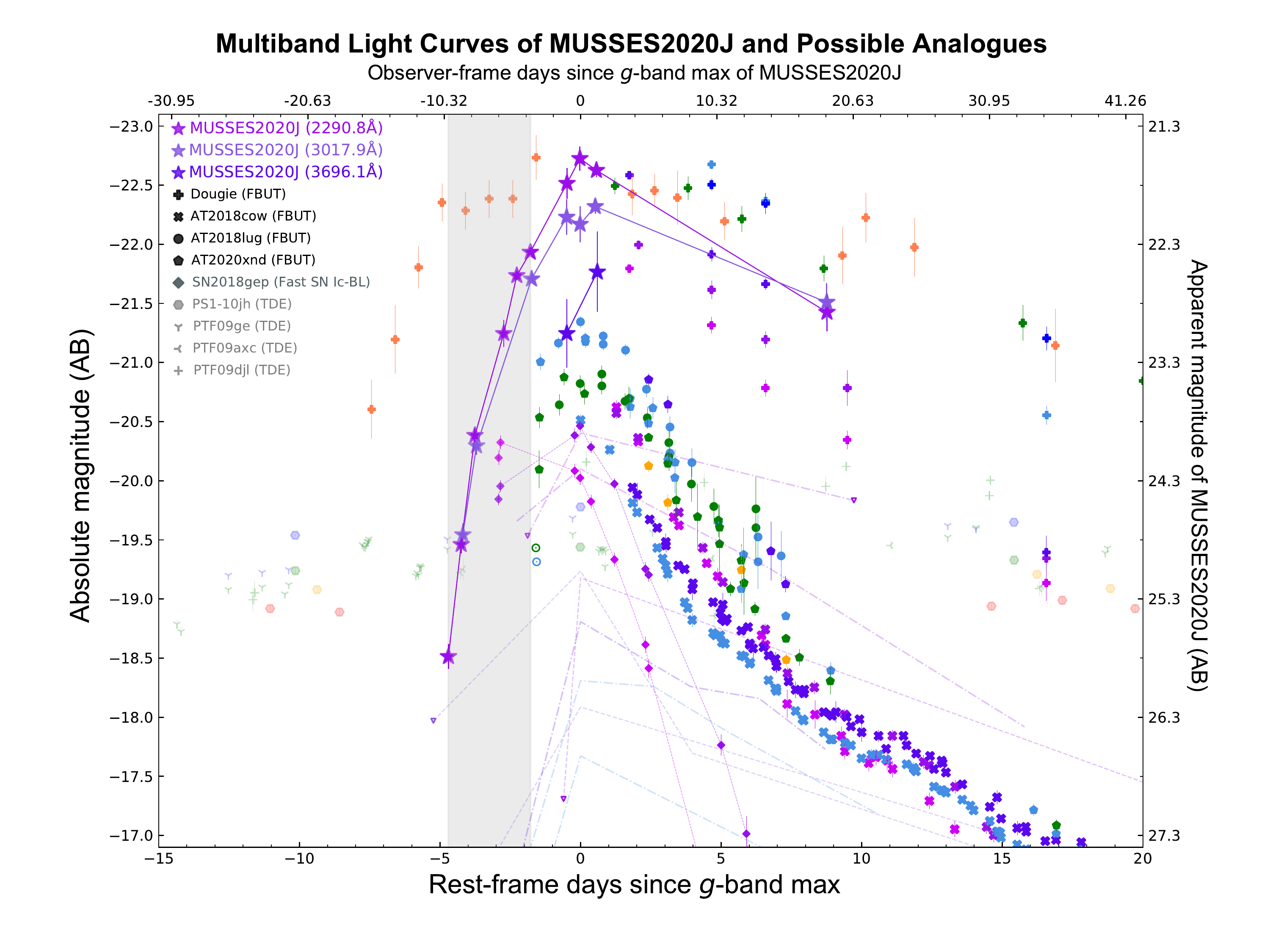}{0.95\textwidth}{}}
\vspace{-30pt}
\caption{Light curves of MUSSES2020J, FBUTs (\citealp{vinko15,perley19,ho20,perley21}; here we classify Dougie as an FBUT due to the extremely high luminosity and fast light-curve decline in UV), SN~2018gep \citep{ho19b}, and normal FBOTs discovered by previous HSC and PS1 observations (broken lines; 3-$\sigma$ $g$-band non-detection limits are given as open triangles; \citealp{tanaka16,tampo20,drout14}). Light curves of some optically selected TDEs and candidates (\citealp{gezari12,arcavi14,blagorodnova17}) are plotted with transparent symbols for comparisons. Galactic extinction of MUSSES2020J has been corrected, assuming an extinction to reddening ratio $R_{V}$ of 3.1 \citep{schlafly11}. The gray stripe indicates the period of the MUSSES2020 Subaru/HSC observation. The 3-$\sigma$ non-detection limits of 27.4 mag ($g$) and 26.2 mag ($r$) of Subaru/HSC observations at about 160 days (rest frame) after the discovery are not shown on the plot. Transients with comparable rest-frame central-wavelength light curves are given by similar colors, and redder colors represent observations in redder rest-frame wavelengths.
\label{fig:FBOT_LCs}}
\end{figure*}

To make the best use of the MUSSES follow-up resources, follow-up observations can be triggered only for transients satisfying specific criteria. Imaging follow-up observations of MUSSES2020J were successfully triggered by the Astrophysical Research Consortium (ARC) 3.5 m telescope and the 3.6 m Devasthal Optical Telescope (DOT) in the next several days after the Subaru observation. However, due to weather and moon phase influences, the MUSSES follow-up network mainly focused on nearby HSC fast transients between late 2020 December and early 2021 January, and we triggered the multiband follow-up observations of MUSSES2020J again on 2021 January 8. Data taken from follow-up observations were reduced in a standard manner for the photometry. Aperture photometry was then performed after image subtraction by matching the scale to the template image. Due to the faintness of MUSSES2020J and limited spectroscopic follow-up time, in addition to a spectroscopy of the host galaxy of MUSSES2020J in 2021 July, no spectroscopic observation was triggered for MUSSES2020J.

In 2021 November, we carried out HSC deep-imaging observations again at the MUSSES2020J field. By stacking HSC $g$-band (2096s on-source time) and $r$-band (515s on-source time) images taken on 2021 November 2--7 (UTC; about 160 days after the discovery in rest frame), we get 3-$\sigma$ non-detection limits of 27.4 mag and 26.2 mag, respectively. All photometric results are given in Table \ref{tab:photometry_data}.

\section{Characteristics of MUSSES2020J}

Figure \ref{fig:FBOT_LCs} presents multiband light curves of MUSSES2020J and some extragalactic transients that may be intrinsically related to MUSSES2020J. A very fast-rising light curve of MUSSES2020J was obtained during the Subaru/HSC survey (gray stripe in the figure), with a $g$-band brightness increase of a factor of about 23 in 5 days (observer frame). Follow-up observations show the brightness of MUSSES2020J kept increasing in the next 4 days after the last Subaru/HSC observation, indicating rise time as short as $\sim$10 days in the observer frame.

Based on the HSC SSP multiband ($grizy$) photometry, physical parameters such as the stellar mass ($M_*$) and star formation rate (SFR) of galaxies in the MUSSES2020 survey fields are estimated from the galaxy template-fitting code MIZUKI \citep{tanaka15,tanaka18}, which performs a fitting with stellar population synthesis templates parameterized by the initial mass function, star formation history, dust attenuation, and metallicity. Given the shortest projected distance and the lowest photometric redshift (also the highest brightness) among all galaxies within 10 arcsec from MUSSES2020J\footnote{According to the photo-$z$ information, the second nearest galaxy, which is a dwarf galaxy with a similar photo-$z$ to that of the nearest one, is already $\gtrsim20$ kpc (about 4 times larger than the size of the galaxy) away from MUSSES2020J.}, we identify the nearest galaxy as the host galaxy of MUSSES2020J. A photometric redshift of $\sim$0.97 $\pm$ 0.13 (68\% confidence level, $\chi^2 = 3.74$) for the host galaxy gives an extremely high brightness and a very fast-rising light curve of MUSSES2020J. 

\begin{figure*}
\gridline{\fig{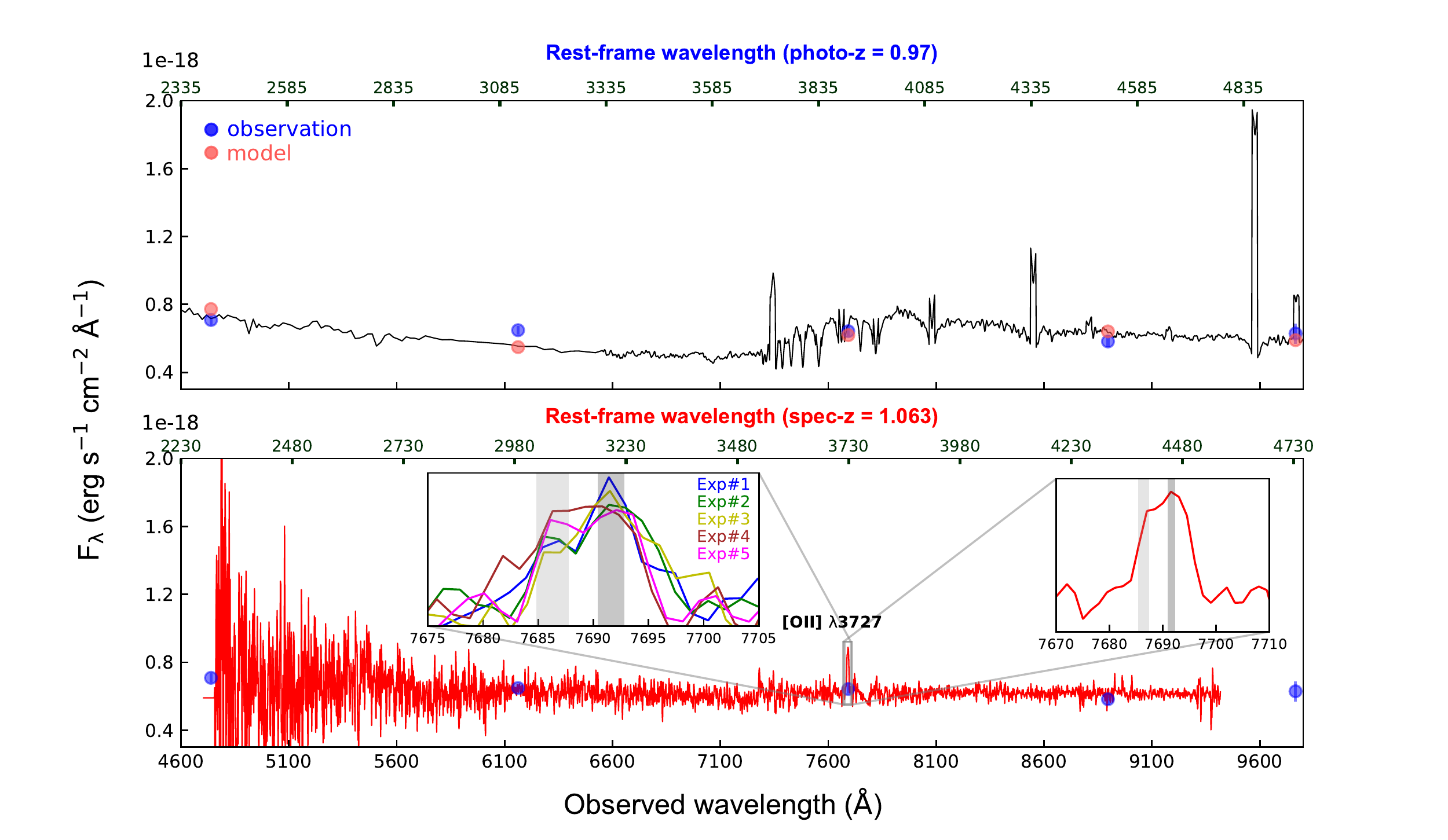}{1.0\textwidth}{}}
\vspace{-25pt}
\caption{Upper panel: A host spectrum modeled by HSC SSP multiband photometries (blue points; \citealp{tanaka15}). Red pointings are synthesized photometries with the model spectrum. Lower panel: A combined 1200 s $\times$ 5 host spectrum taken by Gemini North/GMOS on 2021 July 18 (UTC). A bimodal emission feature around observed wavelength 7690 \AA~(right zoom-in inset) was detected in all 1200s exposures (left zoom-in inset). The central wavelengths of 7686.3 \AA~and 7691.6 \AA~of the gray stripes correspond to the [O {\footnotesize II}] $\lambda$3726.1 and $\lambda$3728.8 lines at a redshift of 1.0628, respectively.
\label{fig:MUSSES2020J_hostspec}}
\end{figure*}

In order to confirm the rest-frame photometric behavior of MUSSES2020J, we triggered a spectroscopic observation of the host galaxy of MUSSES2020J at the Gemini-North observatory, using GMOS \citep{hook04} with the R400+G5305 grating on UT July 18.55, 2021. A total of five exposures were acquired, each with a 1200 s exposure time. Standard CCD processing was accomplished with IRAF. The spectra were extracted using the optimal algorithm of \citealp{horne86}. We fit a low-order polynomial to calibration-lamp spectra and establish the wavelength scale. Small adjustments derived from night-sky lines in the object frames were applied. We further employed our own IDL routines to calibrate the flux and remove telluric lines. Details of our spectroscopic reduction techniques are described by \citet{silverman12}. The photometric redshift fitting and a combined 1200s $\times$ 5 GMOS spectrum of the host galaxy of MUSSES2020J are given in the upper and lower panels of Figure \ref{fig:MUSSES2020J_hostspec}, respectively. The spectrum does not show any AGN features. Notably, a bimodal emission feature that has been detected in all 1200 s exposures around the observed wavelength of $7690$\AA~shows good consistency with the [O {\footnotesize II}] $\lambda$3726.1 and $\lambda$3728.8 lines at a redshift of 1.063. We note that the emission feature cannot be [O {\footnotesize III}] $\lambda\lambda$4959, 5007 as the width of the emission feature of the host at a corresponding redshift should be much narrower than the observed width of [O {\footnotesize III}] $\lambda\lambda$4959, 5007. Given the robust detection of the emission-line doublet, hydrogen emissions such as H$\alpha$ or Ly$\alpha$ can be excluded. By adopting the host redshift of 1.063, MUSSES2020J shows a much higher peak luminosity than that of AT~2018cow at comparable wavelengths (Figure \ref{fig:FBOT_LCs}), breaking the UV luminosity record of FBOTs. On the other hand, multiband follow-up observations in early January 2021 confirmed that the brightness of MUSSES2020J decreased to $\sim$23 mag in both $g$ and $r$ bands, corresponding to a decreasing speed of more than 1.3 mag in 10 days in the rest frame. The light curve of MUSSES2020J evolves much faster than those of superluminous SNe that have comparable peak luminosities and is also much brighter and slower than that of the fast-evolving luminous broad-lined Type Ic SN~2018gep at similar wavelengths (Figure \ref{fig:FBOT_LCs}; \citealp{ho19b}).

In contrast to previously discovered FBUTs whose rise time was roughly estimated based on the last non-detection of the objects, multiband rising-phase light curves obtained from Subaru/HSC and follow-up facilities enable us to pin down the rise time and study the rising-phase behavior of FBUTs for the first time. The deep HSC detection constrains the rise time of MUSSES2020J to about 5--6 days based on the observed peak epoch in the rest-frame UV wavelengths\footnote{The real rise time might be slightly longer than the observation-constrained rise time if there is a ``plateau-like" phase after the observed $g$-band peak, though such a feature has not been seen in UV wavelengths of FBOTs.}, suggesting a much higher energy output than that of AT~2018cow.

\begin{figure*}
\gridline{\fig{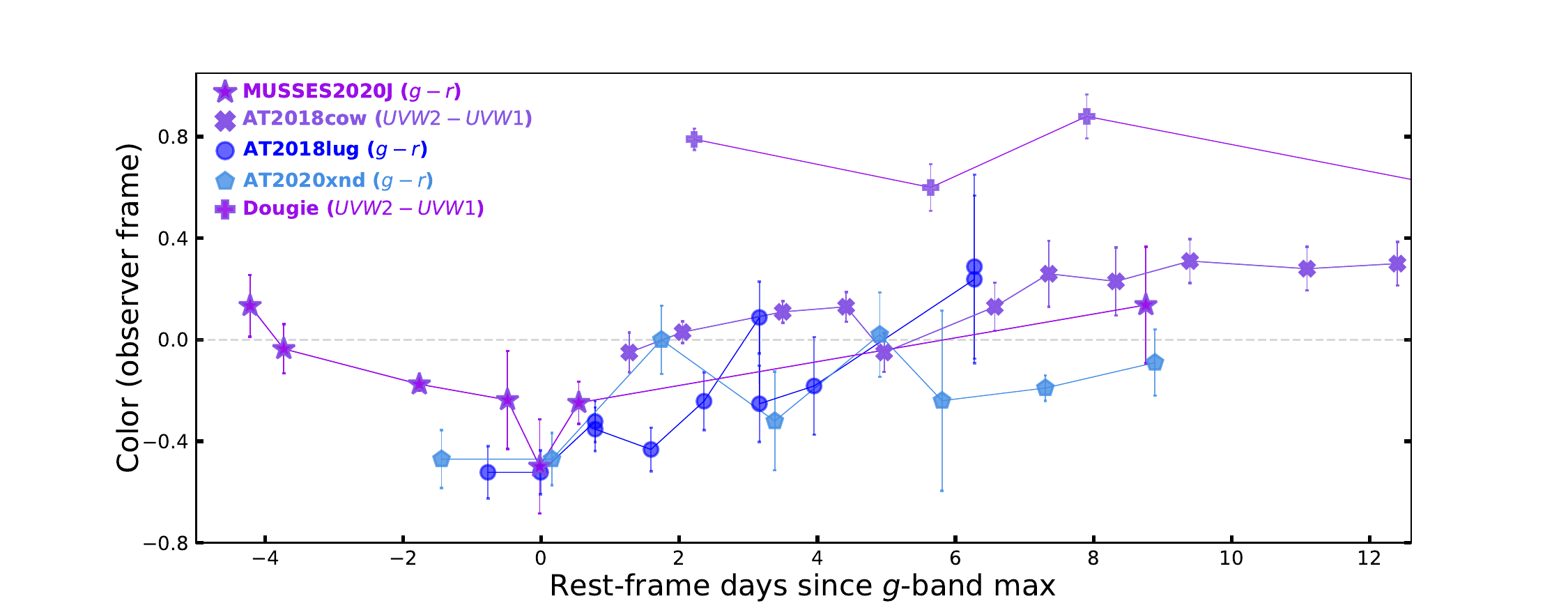}{0.9\textwidth}{}}
\vspace{-15pt}
\caption{Color evolution of MUSSES2020J and other FBUTs. The color evolution of MUSSES2020J shows that the UV color is relatively red in the early time and gets bluer slowly in the light-curve rising phase. Considering the redshift effect, the observed $g-r$ color evolution of MUSSES2020J is consistent with the $UVW2-UVW1$ color of AT~2018cow in the post-maximum phase. In addition to a peculiar plateau feature in the optical light curve of Dougie (Figure \ref{fig:FBOT_LCs}), its post-maximum UV color evolution is redder and slower than those of MUSSES2020J and AT~2018cow, suggesting a lower temperature and possibly a different origin from other FBUTs.
\label{fig:FBUT_color_evo}}
\end{figure*}

Although a large fraction of HSC time was weathered out during the MUSSES2020 campaign, we fortunately observed the color evolution from a very early time, thanks to the fast brightening of MUSSES2020J (Figure \ref{fig:FBUT_color_evo}). The color is overall blue with relatively slow variance, as is consistent with previous samples of FBUTs. It is worth noting that the color evolution of MUSSES2020J changes its behavior around the peak epoch; the color evolves from red to blue in the premaximum phase and gets redder afterward. The early light curve and especially the early color evolution of MUSSES2020J, as obtained for the first time for FBUTs, will be key to constraining the physical model of FBUTs. A detailed discussion of the physical properties derived from the early multiband photometry is given in Section 5. Basic information on FBUTs is summarized in Table \ref{tab:transients_properties}.

\section{The host galaxy and the location of MUSSES2020J}

With the host redshift of 1.063 derived from the Gemini/GMOS spectroscopy, we further investigate the host properties by fixing the spec-$z$ of the host with MIZUKI. The estimated stellar mass and SFR are about $5.65_{-1.87}^{+3.05}\times10^{9}$ $M_\odot$ and $3.65_{-1.13}^{+1.58}$ $M_\odot$/yr (68\% confidence level), respectively. A specific star formation rate (SSFR) of $0.65_{-0.36}^{+0.73}$ Gyr$^{-1}$ also suggests a modest star-forming activity for the host galaxy at a redshift of about 1 \citep{pearson18}. The key properties of the host galaxy are summarized in Table \ref{tab:host_galaxy}.

\begin{figure*}
\gridline{\fig{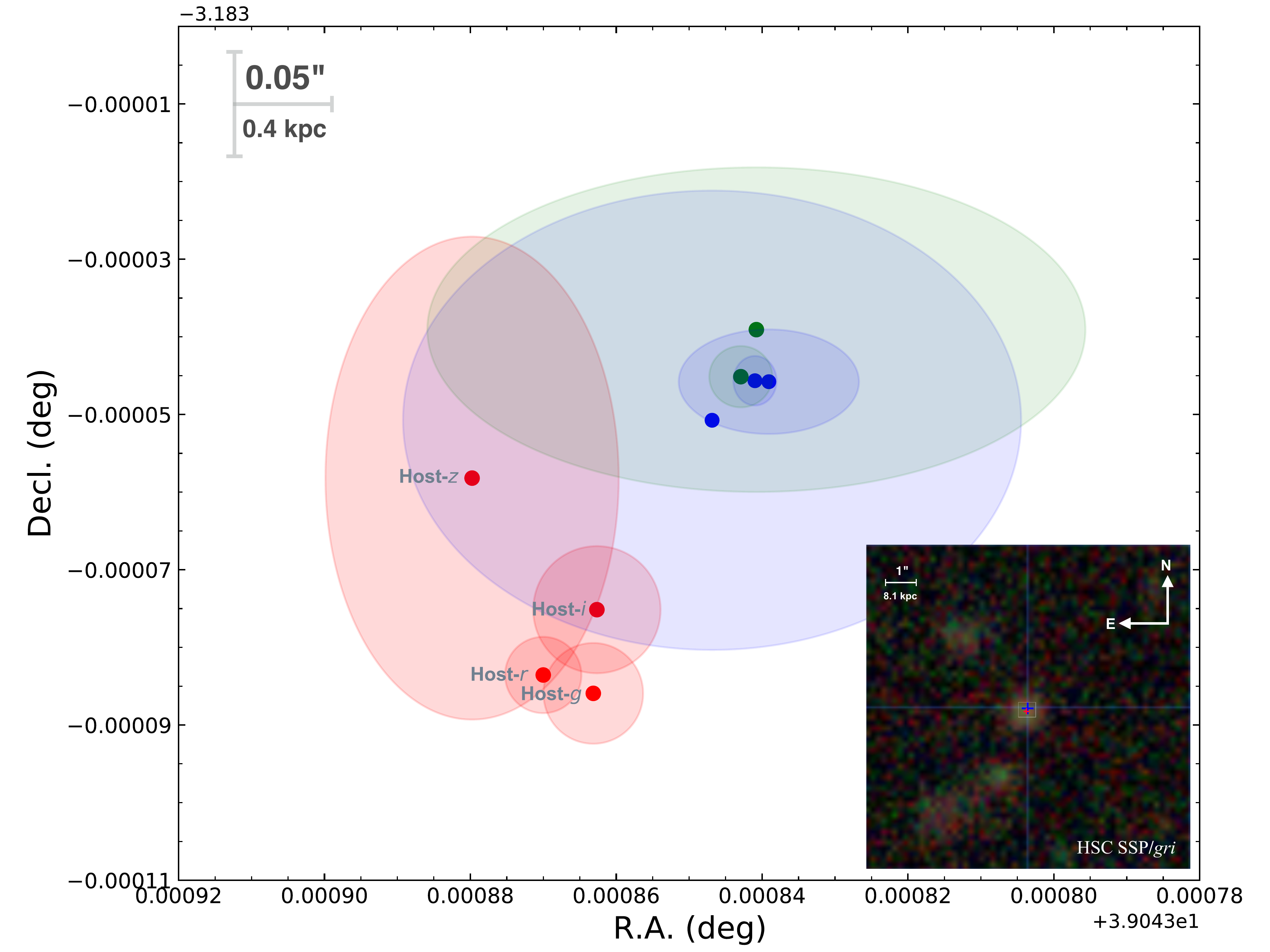}{0.85\textwidth}{}}
\vspace{-20pt}
\caption{Locations of MUSSES2020J and the nucleus of the host galaxy. Red spots are centers of the host galaxy measured from HSC SSP broadband images ($griz$). The coordinates of MUSSES2020J in $g$ and $r$ bands are shown by blue and green spots, respectively. Due to the faintness of MUSSES2020J in the first two nights, here we only present coordinates measured from observations from the third night. 1-$\sigma$ astrometric uncertainty regions of each measurement are denoted by transparent colors. The entire selected area is denoted by a white box at the center of the HSC $g$, $r$, and $i$ composite multicolor thumbnail image of the host galaxy shown in the bottom-right corner. Blue and red crosses in the box indicate the locations of MUSSES2020J ($g$ band) and the host nucleus ($i$ band), respectively. \label{fig:MUSSES2020J_position}}
\end{figure*}

As shown in Figure \ref{fig:MUSSES2020J_position}, MUSSES2020J is located at very close to the center of the host galaxy. Given that the coordinate scatter of MUSSES2020J is mainly on the northwest side of the host nucleus and the 1-$\sigma$ astrometric uncertainty regions of some exposures of MUSSES2020J do not overlap with those of the galaxy center determined by HSC SSP, we judge that MUSSES2020J is a near-nuclear transient located at $\sim$0.12--0.21 arcsec ($\sim$0.97--1.70 kpc at $z$ = 1.063) from the host nucleus. However, due to the astrometric uncertainties of both the galactic nucleus and MUSSES2020J from HSC SSP and MUSSES2020J observations (Figure \ref{fig:MUSSES2020J_position}), which gives a 1-$\sigma$ offset range of 0--0.40 arcsec, we cannot completely rule out that MUSSES2020J is a nuclear transient.

\section{Discussion and Conclusions}
\subsection{Origin(s) of MUSSES2020J and Other FBUTs}

Previous studies have shown that the ultrahigh luminosity and fast-evolving light curves of FBUTs cannot be powered by $^{56}$Ni decay through canonical SN explosion scenarios. Follow-up multiwavelength observations further suggest that compact objects are very likely required to either power FBUTs as central engines \citep{perley19,mohan20,pasham21,uno20a} and/or facilitate X-ray emissions \citep{margutti19,gottlieb22,soker22}. In the following, we discuss possible energy sources of MUSSES2020J suggested by the observational properties of MUSSES2020J and its host galaxy. We leave a systematical theoretical investigation of possible scenarios of MUSSES2020J to a separate paper (K. Uno et al. 2022, in preparation).

\begin{figure*}
\gridline{\fig{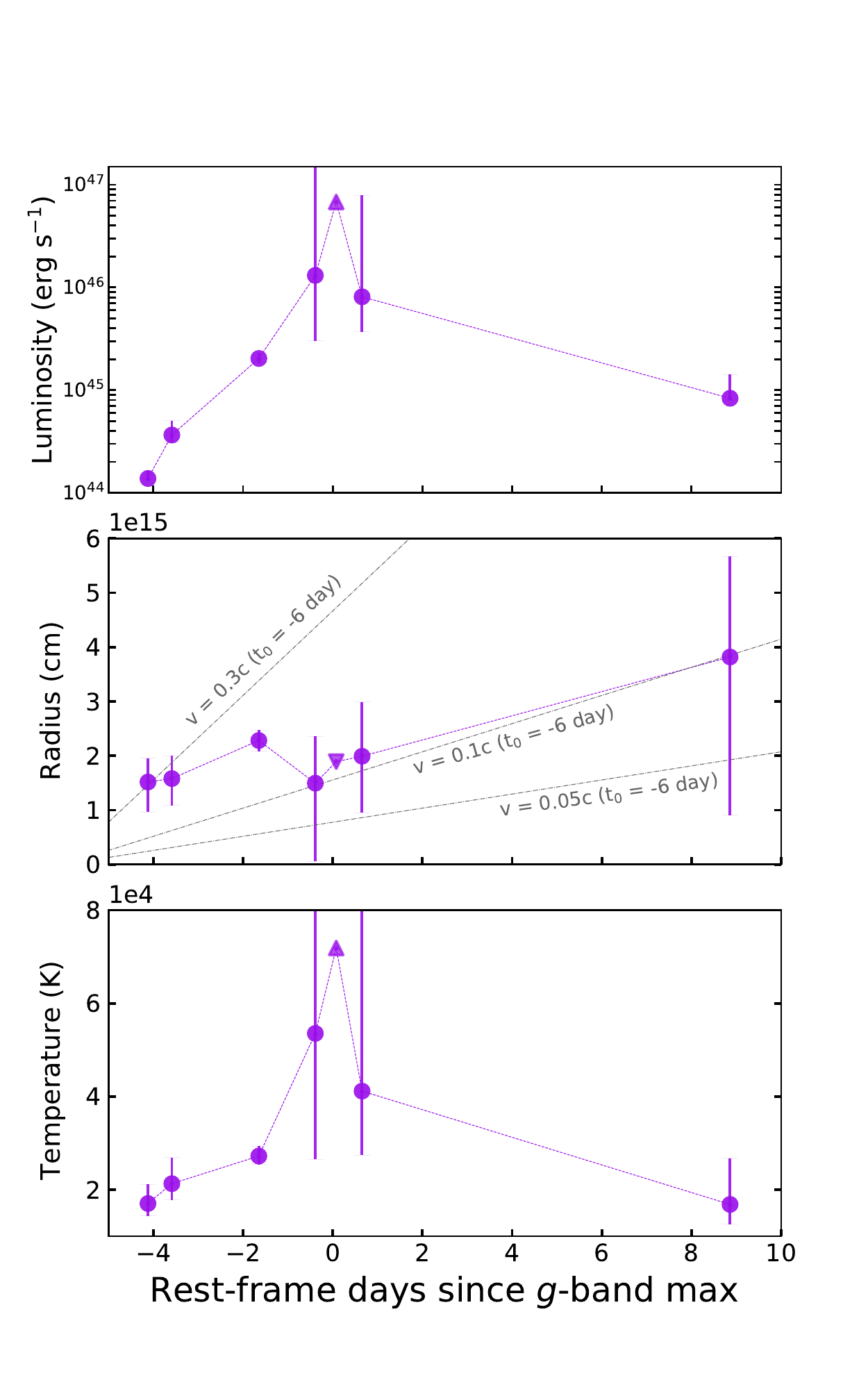}{0.40\textwidth}{}}
\vspace{-38pt}
\caption{Physical properties based on the blackbody fits of multiband observations of MUSSES2020J. In the top panel, data points show the luminosity inferred from the sum of the total Stefan--Boltzmann luminosity of the thermal component. Data points in the middle and bottom panels show the evolution of the best-fit photospheric radius and blackbody temperature, respectively. Error bars of the panels are provided at the 1-$\sigma$ confidence level. Around the $g$-band maximum when the temperature is high (as inferred from the color evolution in Figure \ref{fig:FBUT_color_evo}), the peak of the blackbody radiation curve is at a very short wavelength. Only the lower limits for the blackbody temperature and luminosity, and thus the upper limit for the radius, can be placed by the blackbody fit in the observed wavelength regime. The triangle in each panel corresponds to the 4-$\sigma$ upper/lower limit of each parameter.
\label{fig:BB_para_evo}}
\end{figure*}

By applying a blackbody fit to the multiband light curves of MUSSES2020J, the estimated blackbody radius reveals an interesting behavior (middle panel in Figure \ref{fig:BB_para_evo}). It increases quickly toward the peak luminosity at first. Around the peak, it decreases and then increases again at a slower rate than the initial phase. The initial rise corresponds to an average velocity of $\sim$100,000 km s$^{-1}$ by conservatively assuming a rest-frame time interval of 2 days between the transient occurrence and the second-night HSC observation (i.e., the first multiband observation). In the post-peak phase following the drop, the corresponding velocity is $\sim$30,000 km s$^{-1}$.

The velocity of the photospheric expansion in the post-peak phase has a similarity to what was inferred for AT~2018cow in its earliest phase \citep{perley19}. The behavior is however different, i.e., there is a further increase of the photosphere in MUSSES2020J while there is a decrease in AT~2018cow. This may be related to the slower evolution of MUSSES2020J than AT~2018cow. Even more interesting is the extremely high velocity in the initial phase of MUSSES2020J ($\sim$0.3$c$), which may be related to a sub-relativistic component similar to what was inferred for AT~2018cow from radio emission \citep{ho19a}.

Observationally, although the UV brightness of MUSSES2020J is even brighter than that of AT~2018cow, the slower light-curve evolution makes it possible for the prevailing origin scenarios of FBUTs to explain MUSSES2020J by varying input parameters (e.g., the strength of the magnetic field and spin period of the magnetar model; the mass-loss rate and the position where the wind is launched in the wind-driven scenario). We caution that a strong motivation for interpreting previous FBUTs via SN- or stellar-mass compact object-related scenarios is the possible relation with active star-forming environments. However, the near-nuclear location in the normal low-mass galaxy of MUSSES2020J may suggest that some FBUTs do not rely on the high star-forming environment and might relate to the central part of the host galaxy.\footnote{Note that AT~2020xnd also can be a (near-)nuclear FBUT given the large astrometric uncertainty of its host galaxy \citep{perley21}.}

If the high star-forming environment is not necessary, all FBUTs discovered in low-mass galaxies yield another possibility: an intrinsic connection between FBUTs and IMBHs or low-mass supermassive black holes (SMBHs). Then, IMBH-/SMBH-induced TDEs via the wind-driven scenario could be promising energy sources of FBUTs. Moreover, off-nuclear locations of previous FBUTs cannot rule out the possibility of an IMBH TDE origin as lower-mass IMBHs can also be formed far away from host nuclei (e.g., centers of globular clusters), which indeed is in line with previous estimates of a $\sim$10$^4$ $M_{\odot}$ IMBH TDE as a possible origin for AT~2018cow \citep{perley19,uno20b}. By applying a similar procedure for MUSSES2020J, we can infer a possible range of the BH mass from the best-fit result of the wind-driven TDE scenario (\citealp{uno20a}; K. Uno et al. 2022, in preparation) and conclude that it is consistent with a BH mass of about $1 \times 10^6$ $M_{\odot}$ as expected from the $M_{BH}$--$M_{*}$ correlation derived from nearby early and late-type galaxies (gray shaded region in Figure \ref{fig:FBUT_color_evo} of \citealp{greene20})\footnote{We caution that whether such a correlation is still tenable at $z \sim 1$ is not yet clear.}.

In addition to the wind-driven TDE scenario, prominent UV emission might be associated with a synchrotron emission possibly from a (hidden) relativistic jetted TDE \citep{zauderer11}. Early-phase SEDs of MUSSES2020J derived from the HSC $g$- and $r$-band observations suggest that a UV spectral index $\alpha$ ($F_\nu \propto \nu^{\alpha}$) was initially negative ($\alpha \sim -0.4$ on observation day 2 of HSC observations) and later became positive ($\alpha \sim 0.2$ and 0.7 on observation days 3 and 6, respectively). This is against such a possibility; it is generally expected that the synchrotron flux index changes from positive to negative as the emitting region evolved from optically thick to thin, i.e., opposite to what is observed. We caution that it is still possible that (only) the earliest UV emission is powered by the synchrotron emission from a relativistic jet, while the later and bluer emission is powered by thermal emission, probably from a non-relativistic component.

\subsection{Further Implications from the Early Photometric Evolution of MUSSES2020J}

Even though the wind-driven TDE scenario can reasonably explain the photometric behavior of MUSSES2020J, it is hard to exclude models that also have flexible input parameters (e.g., magnetar central engine, mass accretion with a BH formed by a failed SN) and answer whether all FBUTs discovered so far have the same origin/mechanism, due to the limited constraints from our follow-up observations. For example, a $g$-band non-detection limit of about 200 times fainter than the peak at $t\sim$ 160 days is still too shallow to provide a tight constraint due to the ultrafast evolution of FBUTs. Nevertheless, the red color of the early-phase MUSSES2020J might not have been considered by previously proposed scenarios due to the absence of early color information of other FBUTs. Indeed, the blue-to-red evolution, as opposed to the observed evolution, is generally expected in a system that launches an optically thick material, irrespective of the model details. Given its optically thick nature, the photosphere will be formed at the outer edge of the expanding materials, and the increasing photospheric radius generally leads to a temperature decrease. This is what is seen in the so-called cooling-envelope emission that describes the earliest emission from Type II SNe \citep{nakar10,rabinak11}. The observed red-to-blue evolution thus poses a challenge for theoretical interpretation and inversely may shed light on uncovering the nature of the FBUTs. The magnetar scenario may have the potential to explain such a behavior by realizing a rapid increase in the bolometric luminosity to compensate the photosphere expansion, while this may require fine-tuning of the model parameters. As an alternative scenario, we may consider the situation that the photosphere would not expand substantially, which could be realized in the SN-CSM interaction and wind-driven scenarios; the photosphere is formed within either the CSM or materials ejected before the main outburst.

The high velocity indicated by the earliest-phase photosphere evolution through the blackbody fit (middle panel of Figure \ref{fig:BB_para_evo}) might suggest that the photosphere is formed within a relatively small amount of the high-velocity material ($\sim$0.3$c$) which might have been ejected preceding the main outburst. As for the peculiar evolution of the photospheric radius toward the $g$-band peak, it is hard to judge if the behavior is real, due to the large uncertainty from the blackbody fit\footnote{Also note that blackbody-fit parameters measured from just a few bands are susceptible to any emission/absorption lines that may be present.}. However, if this evolution is real, it might suggest that the high-velocity ejecta become transparent and the position of the photosphere ``jumps" to the inner and slower material (the main ejecta with $\sim$0.1$c$) before the peak. If the inner ejecta are not yet optically thin, the photosphere expands again but with a slower velocity than previously. This could be viewed as a variant of the wind-driven scenario, at least qualitatively. We note that there is a possibility that the earliest emission could be associated with a non-thermal emission from a relativistic jet (Section 5.1), but the above argument applies even in such a case. More comprehensive theoretical investigations will answer whether a specific physical process is necessary to interpret the early photometric evolution of MUSSES2020J, which will provide stringent constraints on the theoretical models of FBUTs (K. Uno et al. 2022, in preparation).

\subsection{Searching for FBUTs in the Era of Deep Wide-field Optical Surveys}

As proved by the higher redshifts (except for AT~2018cow) of previously discovered FBUTs and the early discovery of MUSSES2020J during a short-term HSC survey (a 10  half-night survey with $\sim$33\% completeness), their extremely high brightness in the UV and blue optical wavelengths makes FBUTs promising targets in upcoming deep wide-field optical surveys. In particular, given a more prominent time dilation at higher redshift, the intra-day variability of high-$z$ FBUTs can be systematically investigated by daily cadence deep-imaging surveys. 

In the coming few years, the new-generation deep-imaging optical survey facilities such as the 2.5m Wide Field Survey Telescope (WFST), led by the University of Science and Technology of China (USTC) and Purple Mountain Observatory (PMO), and the Vera C. Rubin LSST will start scientific observations. The excellent site \citep{deng21} and the superior survey capability make WFST the most powerful $u$-band survey facility in the northern hemisphere, and dozens of FBUTs at a wide redshift coverage can be expected from a deep high-cadence $u$-band survey of WFST in a few years. With the dramatically increasing number of FBUTs discovered in the near future, we will be able to not only figure out the origin of this newly confirmed extreme transient class but perhaps also offer a novel approach to investigating the formation and evolution of IMBHs/SMBHs in the era of deep wide-field time-domain surveys.

\subsection{Summary}

In this Letter we report the discovery of an ultraluminous fast transient, MUSSES2020J at redshift 1.063, soon after its occurrence with Subaru/HSC. The ultraluminous fast-evolving light curve and a low-mass host galaxy share similarities with previously discovered fast blue ultraluminous transients (FBUTs) whose peak brightnesses are comparable to or even higher than that of AT~2018cow. A near-nuclear location of MUSSES2020J might imply a possible connection between the energy source of MUSSES2020J and the central part of the host galaxy. The color behavior and physical properties derived from the early multiband light curve suggest that previously proposed scenarios such as the wind-driven scenario and the SN-CSM interaction can at least qualitatively explain the general features of MUSSES2020J. More detailed theoretical work focusing on the multiband light-curve fit is required to answer the origin of this extreme transient. Given their ultrahigh luminosities in the UV and blue optical wavelengths, a promising number of FBUTs from the local to the high-$z$ universe can be discovered through upcoming deep wide-field optical surveys.

\begin{acknowledgments}
We thank the anonymous referee for helpful comments and suggestions. This work has been supported by the World Premier International Research Center Initiative (WPI) and the Japan Society for the Promotion of Science (JSPS) KAKENHI grants JP18J12714, JP19K23456, and JP22K14069 (J.J.), JP16H01087 and JP18H04342 (J.J. and M.D.), JP18H05223 (J.J., K.M., M.D., and T.S.), JP20H00174 and JP20H04737 (K.M.), JP15H02082, JP16H06341, and JP16K05287 (T.S.), JP19H00694, JP20H00158, and JP20H00179 (M.T.), JP17K05382, JP20K04024, and JP21H04499 (K.N.). D.J. acknowledges support from the Erasmus+ programme of the European Union under grant No. 2020-1-CZ01-KA203-078200.

The Hyper Suprime-Cam (HSC) collaboration includes the astronomical communities of Japan and Taiwan, and Princeton University. The HSC instrumentation and software were developed by the National Astronomical Observatory of Japan (NAOJ), the Kavli Institute for the Physics and Mathematics of the Universe (Kavli IPMU), the University of Tokyo, the High Energy Accelerator Research Organization (KEK), the Academia Sinica Institute for Astronomy and Astrophysics in Taiwan (ASIAA), and Princeton University. Funding was contributed by the FIRST program from Japanese Cabinet Office, the Ministry of Education, Culture, Sports, Science and Technology (MEXT), the Japan Society for the Promotion of Science (JSPS), Japan Science and Technology Agency (JST), the Toray Science Foundation, NAOJ, Kavli IPMU, KEK, ASIAA, and Princeton University. 
This paper makes use of software developed for the Large Synoptic Survey Telescope. We thank the LSST Project for making their code available as free software at http://dm.lsst.org

The Pan-STARRS1 Surveys (PS1) have been made possible through contributions of the Institute for Astronomy, the University of Hawaii, the Pan-STARRS Project Office, the Max-Planck Society and its participating institutes, the Max Planck Institute for Astronomy, Heidelberg and the Max Planck Institute for Extraterrestrial Physics, Garching, The Johns Hopkins University, Durham University, the University of Edinburgh, Queen’s University Belfast, the Harvard-Smithsonian Center for Astrophysics, the Las Cumbres Observatory Global Telescope Network Incorporated, the National Central University of Taiwan, the Space Telescope Science Institute, the National Aeronautics and Space Administration under Grant No. NNX08AR22G issued through the Planetary Science Division of the NASA Science Mission Directorate, the National Science Foundation under Grant No. AST-1238877, the University of Maryland, and Eotvos Lorand University (ELTE) and the Los Alamos National Laboratory.

This research is based in part on data collected at the Subaru Telescope and retrieved from the HSC data archive system, which is operated by the Subaru Telescope and Astronomy Data Center at NAOJ.

Based on observations obtained with the Apache Point Observatory 3.5 m telescope, which is owned and operated by the Astrophysical Research Consortium.

Based on observations obtained at the 3.6 m Devasthal Optical Telescope (DOT), which is a National Facility run and managed by Aryabhatta Research Institute of Observational Sciences (ARIES), an autonomous Institute under the Department of Science and Technology, Government of India.

Based on observations (GN-2020B-DD-107, GN-2020B-FT-213, GN-2021A-Q-108) obtained at the international Gemini Observatory, a program of NSF’s NOIRLab, which is managed by the Association of Universities for Research in Astronomy (AURA) under a cooperative agreement with the National Science Foundation on behalf of the Gemini Observatory partnership: the National Science Foundation (United States), National Research Council (Canada), Agencia Nacional de Investigaci\'{o}n y Desarrollo (Chile), Ministerio de Ciencia, Tecnolog\'ia e Innovaci\'{o}n (Argentina), Minist\'{e}rio da Ci\^{e}ncia, Tecnologia, Inova\c{c}\~{o}es e Comunica\c{c}\~{o}es (Brazil), and Korea Astronomy and Space Science Institute (Republic of Korea). The observations were carried out within the framework of Subaru-Keck/Subaru-Gemini time exchange program which is operated by the National Astronomical Observatory of Japan. We are honored and grateful for the opportunity of observing the Universe from Maunakea, which has the cultural, historical and natural significance in Hawaii.

\end{acknowledgments}

\facilities{Subaru, Gemini North, APO:3.5 m, DOT:3.6 m.}

\software{hscPipe, IRAF, Astropy \citep{astropy13,astropy18}.}

\bibliography{Bibliography_MUSSES2020J_Jiang.bib}

\begin{deluxetable*}{ccccccc}
\movetabledown=3000mm
\tablenum{1}
\tablecaption{Observations of MUSSES2020J\label{tab:photometry_data}}
\tablewidth{0pt}
\tablehead{
\colhead{UT Date} & \colhead{MJD} & \colhead{Phase$^*$} & \colhead{$g$} & \colhead{$r$} & \colhead{$i$} & \colhead{Telescope/Instrument}
}
\startdata
2020/12/11 & 59194.31 & -4.70 & 25.92 (10) & -- & -- & Subaru/HSC \\
2020/12/12 & 59195.25 & -4.24 & 24.97 (08) & -- & -- & Subaru/HSC \\
2020/12/12 & 59195.36 & -4.19 & -- & 24.84 (09) & -- & Subaru/HSC \\
2020/12/13 & 59196.25 & -3.76 & 24.05 (06) & -- & -- & Subaru/HSC \\
2020/12/13 & 59196.37 & -3.70 & -- & 24.08 (08) & -- & Subaru/HSC \\
2020/12/15 & 59198.36 & -2.73 & 23.18 (12) & -- & -- & Subaru/HSC \\
2020/12/16 & 59199.32 & -2.27 & 22.69 (04) & -- & -- & Subaru/HSC \\
2020/12/17 & 59200.31 & -1.79 & 22.50 (01) & -- & -- & Subaru/HSC \\
2020/12/17 & 59200.40 & -1.75 & -- & 22.67 (02) & -- & Subaru/HSC \\
2020/12/19 & 59202.98 & -0.49 & -- & 22.15 (15) & -- & DOT/ADFOSC \\
2020/12/19 & 59202.99 & -0.49 & -- & -- & 23.10 (29) & DOT/ADFOSC \\
2020/12/20 & 59203.00 & -0.48 & 21.91 (13) & -- & -- & DOT/ADFOSC \\
2020/12/20 & 59203.95 & -0.02 & 21.71 (10) & -- & -- & DOT/ADFOSC \\
2020/12/20 & 59203.96 & -0.02 & -- & 22.21 (15) & -- & DOT/ADFOSC \\
2020/12/22 & 59205.08 & 0.52 & -- & 22.05 (06) & -- & ARC/ARCTIC \\
2020/12/22 & 59205.16 & 0.56 & 21.81 (06) & -- & -- & ARC/ARCTIC \\
2020/12/22 & 59205.23 & 0.60 & -- & -- & 22.58 (34) & ARC/ARCTIC \\
2021/01/08 & 59222.06 & 8.75 & -- & 22.87 (16) & -- & ARC/ARCTIC \\
2021/01/08 & 59222.08 & 8.76 & 23.00 (16) & -- & -- & ARC/ARCTIC \\
2021/11/04 & 59522.5 & 154.4 & 27.40$^{**}$ & -- & -- & Subaru/HSC \\
2021/11/05 & 59523.0 & 154.6 & -- & 26.22$^{**}$ & -- & Subaru/HSC \\
\hline
\enddata
\tablecomments{\\
Magnitudes are in the AB system. Numbers in parentheses correspond to 1-$\sigma$ statistical uncertainties in units of 1/100 mag.\\
$^*$ Days (rest frame) relative to an estimated $g$-max time (MJD = 59204).\\
$^{**}$ 3-$\sigma$ non-detection limits with stacked HSC $g$-band (2096s on-source time) and $r$-band (515s on-source time) images taken on UT 2021/11/02--2021/11/07 and 2021/11/03--2021/11/07, respectively.
}
\end{deluxetable*}

\begin{deluxetable*}{ccccccccc}
\movetabledown=3000mm
\tablenum{2}
\tablecaption{Properties of FBUT and Candidates \label{tab:transients_properties}}
\tablewidth{0pt}
\tablehead{
\colhead{Name} & \colhead{$\alpha$ (J2000)} & \colhead{$\delta$ (J2000)} & \colhead{Redshift} & \colhead{Rise time$^{*}$} & \colhead{Peak Mag$^{**}$} & \colhead{Offset [arcsec]} & \colhead{Offset [kpc]} & \colhead{Ref$^{***}$}
}
\startdata
MUSSES2020J & 02:36:10.52 & $-$03:10:59.0 & 1.063 & 5--6 days (2291\AA) & $-22.72 \pm 0.10 $ ($g$) & 0.12--0.21$^{\dagger}$ & 0.97--1.70 & This paper \\
AT~2018cow & 16:16:00.22 & +22:16:04.9 & 0.014 & $\lesssim$3 days (4704\AA) & $-20.80 \pm 0.05$ ($g$) & 5.9 & 1.7 & (1), (2) \\
AT~2018lug & 02:00:15.19 & +16:47:57.3 & 0.271 & $>$2 days (3820\AA) & $-21.34 \pm 0.05$ ($g$) & $0.28_{-0.13}^{+0.13}$ & $1.16_{-0.54}^{+0.54}$ & (3) \\
AT~2020xnd & 22:20:02.014 & $-$02:50:25.35 & 0.243 & $>$5 days? (3904\AA) & $-21.21 \pm 0.05$ ($g$) & -- & -- & (4) \\
Dougie & 12:08:47.87 & +43:01:20.1 & 0.191 & $\gtrsim$8 days (5211\AA) & $-22.76 \pm 0.19$ ($R$) & 1.3 & 4.1 & (5) \\
\hline
\enddata
\tablecomments{\\
$^{*}$ Rest-frame rise time of light curves at specific optical bands. Corresponding rest-frame central wavelengths are given in parentheses. \\
$^{**}$ Galactic extinction-corrected \citep{schlafly11} absolute magnitudes in specific observed optical bands. \\
$^{***}$ (1) \citet{perley19}; (2) \citet{bietenholz18}; (3) \citet{ho20}; (4) \citet{perley21}; (5) \citet{vinko15}. \\
$^{\dagger}$ A 1-$\sigma$ offset range is $\sim$0--0.4 arcsec considering the astrometric uncertainties of both the host nucleus and the transient.
}
\end{deluxetable*}

\begin{table}
\tablenum{3}
\caption{Properties of the MUSSES2020J Host Galaxy \label{tab:host_galaxy}
}
\begin{tabular}{lc}
\hline
\hline
HSC-$g$ & 24.69 (04) \\
HSC-$r$ & 24.17 (04) \\
HSC-$i$ & 23.70 (05) \\
HSC-$z$ & 23.45 (08) \\
HSC-$y$ & 23.11 (11) \\
\hline
Phot-$z$ & 0.97$\pm$0.13 \\
Spec-$z$ & 1.0628$\pm$0.0003 \\
Luminosity distance (Mpc) & 7125.6 \\
\hline
$M_{*}$ ($M_\odot$) & $5.65_{-1.87}^{+3.05}\times10^{9}$ \\
SFR ($M_\odot$ yr$^{-1}$) & $3.65_{-1.13}^{+1.58}$ \\
SSFR (Gyr$^{-1}$) & $0.65_{-0.36}^{+0.73}$ \\
\hline
\end{tabular}
\end{table}

\end{document}